\documentclass[aps,pra,twocolumn,superscriptaddress]{revtex4-1}

\usepackage{textcomp}
\usepackage{amsmath}
\usepackage{amssymb}
\usepackage{graphicx}
\usepackage{color}
\usepackage{dsfont}
\usepackage{bbm}
\usepackage{wrapfig}

\begin{document}

\title{Realizing topological relativistic dynamics with slow light polaritons at room temperature}

\author{Mehdi Namazi}
\author{Bertus Jordaan}
\affiliation{Department of Physics and Astronomy, Stony Brook University, New York 11794-3800, USA}
\author{Changsuk Noh}
\affiliation{Korea Institute for Advanced Study, 85 Hoegiro, Seoul 1307}
\author{Dimitris G. Angelakis}
\email{dimitris.angelakis@gmail.com}
\affiliation{Centre for Quantum Technologies, National University of Singapore, 3 Science Drive 2, Singapore 1175}
\affiliation{School of Electrical and Computer Engineering, Technical University of Crete, Chania, 73100, Greece}
\author{Eden Figueroa}
\email{eden.figueroa@stonybrook.edu}
\affiliation{Department of Physics and Astronomy, Stony Brook University, New York 11794-3800, USA}

\begin{abstract}
Here we use a slow light quantum light-matter interface at room temperature to implement an analog simulator of complex relativistic and topological physics. We have realized the famous Jackiw-Rebbi model (JR), the celebrated first example where relativity meets topology. Our system is based upon interacting dark state polaritons (DSP's) created by storing light in a rubidium vapor using a dual-tripod atomic system. The DSP's temporal evolution emulates the physics of Dirac spinors and is engineered to follow the JR regime by using a linear magnetic field gradient. We also probe the obtained topologically protected zero-energy mode by analyzing the time correlations between the spinor components. Our implementation paves the way towards quantum simulation of more complex phenomena involving many quantum relativistic particles.
\end{abstract}

\maketitle

\begin{figure*}
\centerline{\includegraphics[width=1.8\columnwidth]{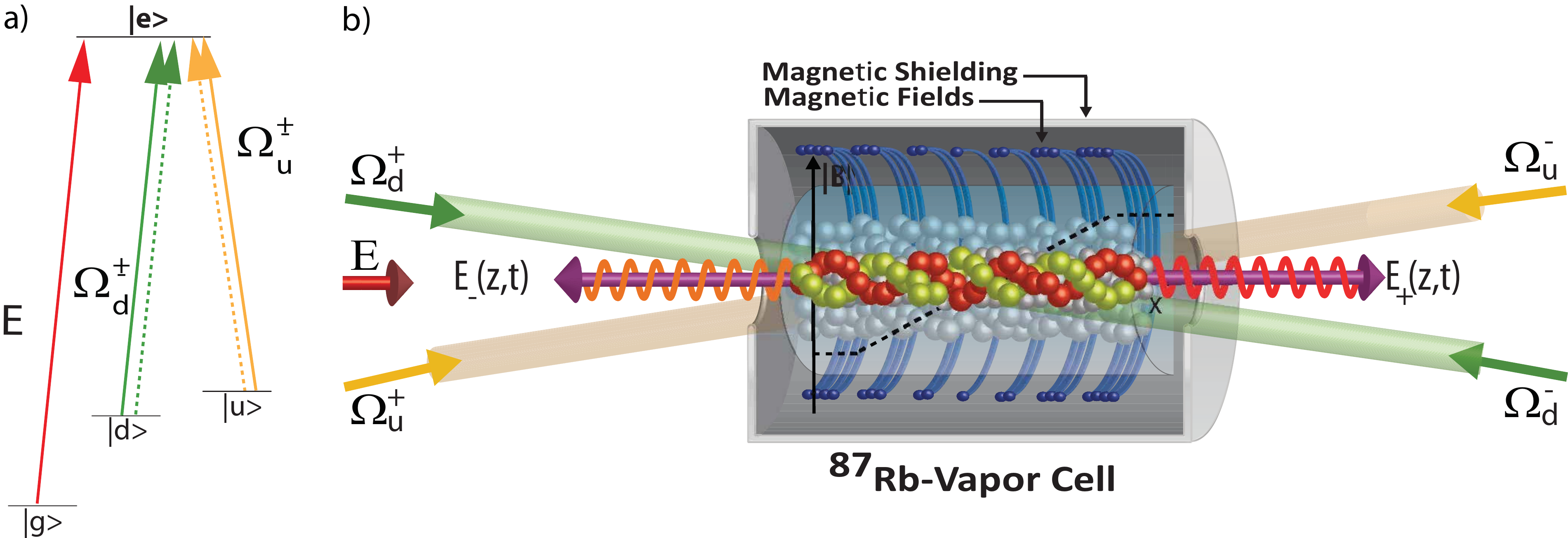}}
\caption{\textbf{Experimental setup for creation of polariton relativistic dynamics at room temperature } (a) The scheme used for creating an EIT tripod system (solid lines) and a dual tripod configuration (solid and dashed lines). (b) In order to experimentally create an EIT tripod system, an electric field $E(z,t)$ (red arrow) enters the medium, simultaneously with two strong co-propagating fields ($\Omega_{d}^{+}$ and $\Omega_{u}^{+})$. For the creation of the dual tripod configuration, the original field $E(z,t)$ is converted into two counter-propagating fields ($E_{-}(z,t)$ and $E_{+}(z,t)$) by introducing two pairs of counter-propagating control fields ($\Omega_{d}^{+}$, $\Omega_{u}^{+}$, $\Omega_{d}^{-}$ and $\Omega_{u}^{-})$. $m_{eff}(z)$ in the JR model is created using a spatially varying magnetic field gradient (dashed black lines).}
\end{figure*}

Over the last decade, a variety of exotic physical phenomena have been realized in artificially created quantum systems near zero temperature \cite{Cirac2012}, including ultra-cold atoms \cite{Bloch2012}, trapped-ions \cite{Blatt2012} and superconducting qubits \cite{Houck2012}. Photonic setups have also been used to emulate relativistic and topological models \cite{Aspuru2012,Rechtsman2013}. An unexplored direction for analog simulation, allowing for operation at room temperature conditions, is the use of atoms interfaced with light in the form of collective excitations known as dark state polaritons (DSPs) \cite{FleischhauerLukin2000}. DSPs have formed the basis of many quantum technology applications including quantum memories \cite{Lvovsky2009,Bussieres2013} and quantum non-linear frequency convertors \cite{radnaev_quantum_2010}. In this work we use DSPs to experimentally demonstrate an analog simulation of the Jackiw-Rebbi (JR) model, the celebrated first example where relativity met topology \cite{JackiwRebbi1976}. Our DSPs are created by storing light in Rb atoms prepared in a dual tripod configuration using counter propagating laser fields. We first show how to realize relativistic Dirac spinor dynamics, as initially suggested in \cite{Unanyan2010}, and then create a static soliton background field as required in the JR model, via a spatially varying atom-photon detuning. We observe signatures of JR's famous zero-energy mode using a temporal analysis of the retrieved light pulses.

DSP's are created by storing and manipulating light in atomic media using electromagnetically-induced-transparency (EIT) \cite{Liu_Nature_2001,phillips_storage_2001,FleischhauerLukin2000,EITReview2005}. In a DSP-based analog simulator, the DSPs are made to follow the dynamics outlined by a light-matter Hamiltonian, prepared by addressing specific atomic transitions using control light fields \cite{Chang2017}. The simulation results are then obtained by measuring the output photon wave functions.  Along these lines, a spinor-like object consisting of two DSPs has been experimentally implemented  in a double tripod configuration \cite{lee_experimental_2014}.

Moreover DSP-based quantum simulators have been proposed to realize Dirac models \cite{Unanyan2010}, and interacting relativistic systems \cite{Angelakis2013}. Among those, the JR model is of paramount relevance \cite{JackiwRebbi1976} as it predicts charge fractionalization \cite{Niemi1986}. This important aspect has been addressed in proposals using optical lattice setups \cite{RuostekoskiJavanainen2002,JavanainenRuostekoski2003}. Additionally, the JR model has gained further attention due to the topological nature \cite{Witthaut2011,Salger2011,Leder2016,Lamata2011,Grossert2016,Casanova2010,Gerritsma2011,Wilczek2016,Zhang2016} of its zero-energy solution \cite{Mugel2016,Tan_PhotTop2014}. This protected mode can be understood as a precursor to topological insulators \cite{Iadecola2016}, a hotly pursued topic nowadays \cite{TIreview1,TIreview2}. Recently, a soliton following a similar model has been observed experimentally in a fermionic superfluid \cite{YefsahZwierlein2013}.

Here we report the experimental realization of relativistic dynamics exhibiting topological aspects as originally conceived in the JR model \cite{JackiwRebbi1976}. We use a room temperature atomic vapour addressed by laser fields in an EIT dual tripod configuration in order to produce slow light Dirac spinor dynamics. We then tune the system to the JR regime by using a linear magnetic field gradient, inducing a kink profile in the corresponding mass term \cite{Angelakis2014}. The topological aspects of this novel EIT light-matter JR system are explored by analyzing the time correlations between the retrieved spinor components.

The structure of the article is as follows: firstly, we review the basics of coherent light propagation in a dual tripod EIT system. We then show the necessary conditions to connect this framework to Dirac and JR dynamics. Lastly, we present the experimental road map to prepare, evolve and benchmark this EIT light-matter JR system in a room temperature atomic interface.
\section{Theoretical Background}
\subsection{Tripod based dark state polaritons.} We start by describing the dynamics of an EIT tripod system (defined as tripod-type linkage pattern in \cite{Ruseckas_2011}) formed by two control fields ($\Omega_{d}$ and $\Omega_{u}$) and a probe field $E(z,t)$   (see Fig. 1a solid lines for definition of atomic levels). Based on the usual EIT assumptions, the following equation describes the propagation of the probe $E(z,t)$ under $\Omega_{d}$:
\begin{equation*}
(\partial_{t}+v_{g}\partial_{z})E(z,t)=+i\frac{g_{\varepsilon}^{2}}{\Omega_{d}^{2}}N \frac{\delta}{2} E(z,t)
\end{equation*}
where $v_{g}=\frac{c}{1+\frac{g_{\varepsilon}^{2}}{\Omega_{d}^{2}}N}$ is the group velocity of the input field in the atomic medium, N is the number of atoms along the beam path, $\delta$ is the two-photon detuning and $g_{\varepsilon}$ is the light-matter coupling constant for $E(z,t)$. In the perturbative and adiabatic limit a similar equation for the atomic operator $\sigma_{gd}$ can be found. The solution to the combined system of equations is a superposition of  $E(z,t)$ and $\sigma_{gd}(z,t)$ and is called a dark state polariton, $\Psi_{d}(z,t)$ \cite{FleischhauerLukin2000}.

In the tripod configuration, the second control field $\Omega_{u}$ creates an additional DSP, $\Psi_{u}(z,t)$. The response of the system is then given by a linear combination of the two DSPs also known as tripod DSPs,$\Psi_{T} =\alpha\Psi_{d}+\beta\Psi_{u}$, with
$\Psi_{d(u)}(z,t) =\cos\theta_{d(u)}  E(z,t)-\sin\theta_{d(u)}\sigma_{gd(u)}(z,t)$ \cite{karpa_resonance_2008}.

\subsection{JR dynamics with DSPs}
Assuming  two pairs of counterpropagating control fields,  see Fig. 1a solid and dashed lines, the evolution of the two probe fields $ E^{+}(z,t)$ and $ E^{-}(z,t)$ can be derived as \cite{Ruseckas_2011}:
 \begin{equation*}
 \label{eqSSL_Omega}
 (\partial_{t}-v_{g}\sigma_{z}\partial_{z})\left(\begin{array}{c}
 E^{+}(z,t)\\
 E^{-}(z,t)
 \end{array}\right)=i\frac{g_{\varepsilon}^{2}}{\bf{\Omega^{2}}}N\sigma_{z}\frac{\delta}{2} \left(\begin{array}{c}
 E^{+}(z,t)\\
 E^{-}(z,t)
 \end{array}\right)
 \end{equation*}

where  $\mathbf{{\Omega}}=\left(\begin{array}{cc} \Omega_{d}^{+} & \Omega_{u}^{+}\\ \Omega_{d}^{-} & \Omega_{u}^{-} \end{array}\right)$.
By individually manipulating the parameters of the control fields, $\mathbf{\Omega}=\Omega\left(\begin{array}{cc}
1 & i\\
i & 1
\end{array}\right)=\Omega(1+i\sigma_{x})$ which results in
\begin{equation}
\label{eqMatrixSSL}
i\hbar(\partial_{t}-v_{g}\sigma_{z}\partial_{z})\left(\begin{array}{c}
E^{+}(z,t)\\
E^{-}(z,t)
\end{array}\right)=\hbar\frac{g_{\varepsilon}^{2}}{2\Omega^{2}}N\sigma_{y}\frac{\delta}{2} \left(\begin{array}{c}
E^{+}(z,t)\\
E^{-}(z,t)
\end{array}\right).
\end{equation}
 It is possible to derive a similar equation for the atomic operators $\sigma^{\pm}(z,t)=\frac{1}{\sqrt{2}}(\sigma_{gu}\pm i\sigma_{gd})$, thus constructing an equation for spinor of slow light (SSL) object $\mathbf{\Psi}=\binom{\Psi^{+}}{\Psi^{-}}$ as:
 \begin{equation}
 \label{eqDirac}
 i\hbar \partial_t \mathbf{\Psi} = \left( i\hbar v_g\sigma_z\partial_z + m_{eff_{0}}v_g^2 \sigma_y \right)\mathbf{\Psi}
 \end{equation}

  with $\Psi^{\pm}(z,t)=\cos\theta E^{\pm}(z,t)-\sin\theta\sigma^{\pm}(z,t)$ and $\theta = \arctan(\sqrt{\frac{g^2N}{\Omega^2}})$ \cite{Unanyan2010}. Equation \ref{eqDirac} resembles a 1+1 Dirac equation and describes the evolution of two coupled SSL components with an effective mass $m_{eff_{0}}=\hbar \frac{\delta}{2} \frac{1}{v_g^2} \sin^2(\theta)$.\\

The JR model, can be realized by adding an extra coupling of this Dirac SSL to a background bosonic field
 \begin{equation}
 i\hbar \partial_t \mathbf{\Psi} = \left( i\hbar v_g\sigma_z\partial_z + m_{eff}(z) v_g^2 \sigma_y \right)\mathbf{\Psi}
 \end{equation}

 where $m_{eff}(z)$ obeys the Klein-Gordon equation. In the case when $m_{eff}(z)$ has a kink profile ($\propto \tanh(z)$), JR showed that it is possible to find a zero-mode solution which is topologically protected:
\begin{equation}
\psi_{zero} = \exp\left(-v_g \int_0^z ds \,m_{eff}(s) \right) \chi,
\end{equation}
where $-\sigma_z \sigma_y \chi=-i\chi $. By denoting $\phi(z)=\exp\left(-v_g \int_0^z ds \,m_{eff}(s) \right) $  and $\chi\propto(1,-1)$, we can write $\psi_{zero}(z) = \phi(z)\begin{pmatrix}1 \\ -1 \end{pmatrix}$ (with suitable normalization). Any particle initialised in this mode will not evolve and stay  localized whereas in the normal Dirac case, it will spread \cite{Angelakis2014}.

\begin{figure}
\centerline{\includegraphics[width=1.0\columnwidth]{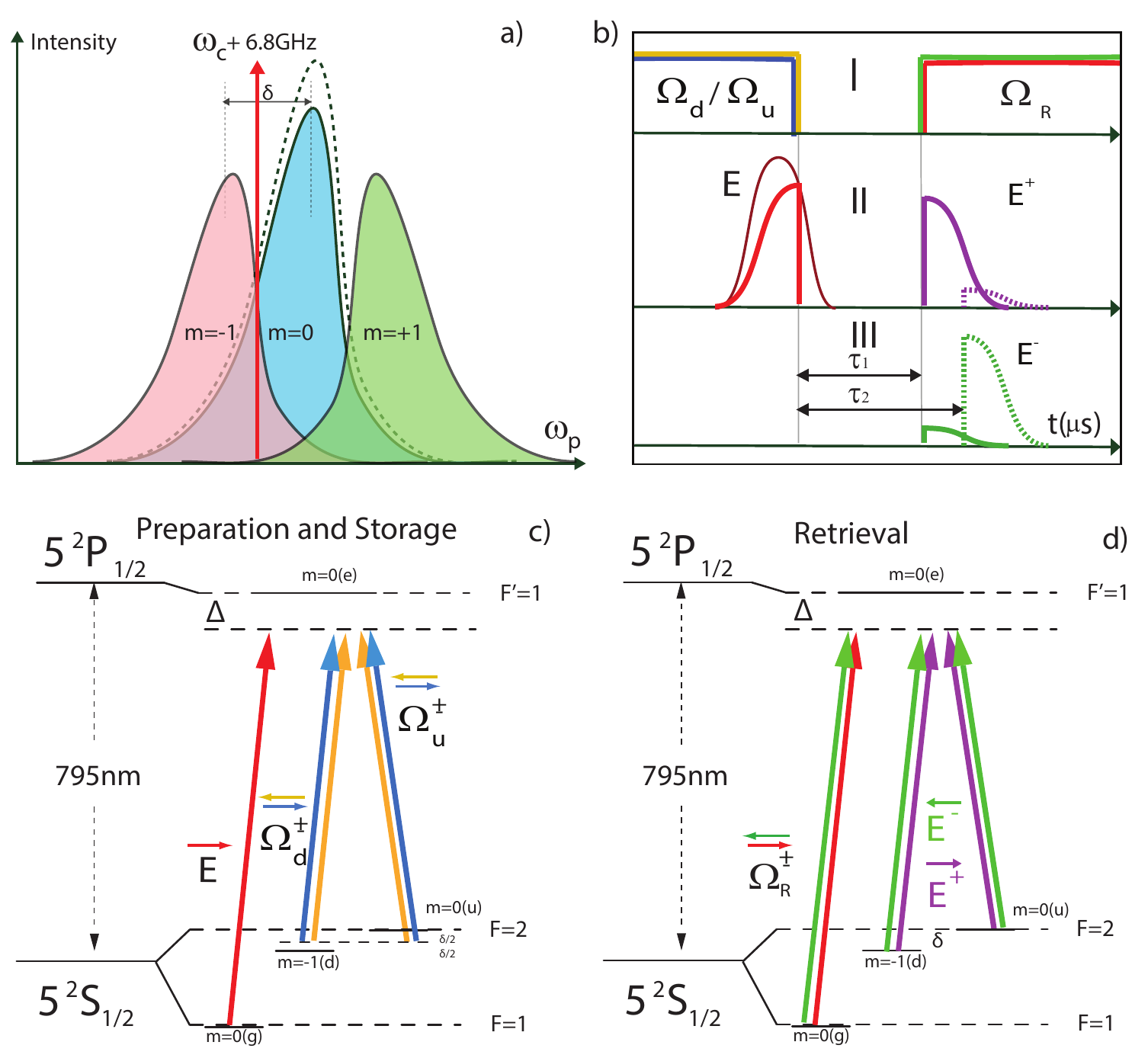}}
\caption{\textbf{Tripod DSP setup.} (a) An original EIT line (dashed black line) is separated into three lines via applying a magnetic field B. Placing the control laser with a proper two-photon detuning ($\pm \delta /2$ from Zeeman states m=0 and m=-1) creates two isolated EIT systems. (b) Pulsing sequence for the creation of dual-tripod dynamics. (c) The SSL $\mathbf{\Psi}$ is created using $\Omega_{u}^{\pm}$ and $\Omega_{d}^{\pm}$. (d) $\mathbf{\Psi}$ is stored for a time $\tau$ after which it is mapped onto $E^{\pm '}$ using $\Omega_{R}^{\pm}$}
\label{FigSpinor}
\end{figure}

\section{Experimental Realization.}
\subsection{Creation of tripod DSP.}
We create the tripod DSP $\Psi_T$ in an atomic ensemble using EIT in the following way. Firstly, three separated EIT systems are created by breaking the degeneracy of the Rb atoms Zeeman sub-levels through applying a DC magnetic field B (see Fig. 2a). Secondly, we isolate two of the EIT systems by using a single control laser that is symmetrically detuned ($\pm \delta/2= \pm g_{d}\mu_{B}B/2$) from the transitions $|u\rangle \rightarrow |e\rangle$ and $|d\rangle \rightarrow |e\rangle$, effectively forming two control fields $\Omega_{u}$ and $\Omega_{d}$ (see Fig. 2b). Lastly, we send a pulse of light ($E(z,t)$) undergoing tripod DSP dynamics due to $\Omega_{u}$ and $\Omega_{d}$, thus creating the components of $\Psi_T$ ($\Psi_u$ and $\Psi_d$).

All the transitions used in the experiment are within the $^{87}$Rb $D_{1}$ line. The storage is based on EIT in warm $^{87}$Rb vapor. The probe is stabilized using top-of-fringe locking to saturation spectroscopy of a Rb vapor cell and the control field is stabilized by an OPLL to the probe field. The probe pulses $E(z,t)$ with a width of 400 $\mu$s is tuned to $5S_{1/2} |F, m_{F}= 1,0\rangle$ $\rightarrow$ $5P_{1/2} |F', m_{F'}= 1,0\rangle$ ($|g\rangle$ $\rightarrow$ $|e\rangle$) (with detuning $\Delta=250MHz$). The writing control fields ($\Omega_{u}$ and $\Omega_{d}$) are tuned at $|F, m_{F}= 2,0\rangle$ $\rightarrow$ $|F', m_{F'}= 1,0\rangle$ ($|u\rangle$ $\rightarrow$ $|e\rangle$ with detuning $-\delta/2$) and $|F, m_{F}= 2,-1\rangle$ $\rightarrow$ $|F', m_{F'}= 1,0\rangle$ ($|d\rangle$ $\rightarrow$ $|e\rangle$ with detuning $+\delta/2$)(see Fig. 2a). The EIT lines have an average FWHM of 1.2 MHz \cite{Namazi2017}. A constant magnetic field induces the two-photon detuning with a $\delta/B$ ratio of 1.09 MHz/G.

The time sequence of the creation of $\Psi_T$ and the readout of $\Psi_{T}^{'}$ is shown in fig. 2b. $\Psi_T$ is stored for $2\mu s$ after which it is mapped onto $E_{d}^{'}$ and $E_{u}^{'}$ using $\Omega_R$.  $\Omega_{R}$ is tuned to $|F, m_{F}= 1,0\rangle$ $\rightarrow$ $|F', m_{F'}= 1,0\rangle$ ($|g\rangle$ $\rightarrow$ $|e\rangle$). We calibrate the coherence of this tripod scheme by storing $\Psi_T$ and retrieving it using a co-propagating control field ($\Omega_{R}$) coupled to the $|g\rangle \rightarrow |e\rangle$ transition. Polarization elements supply 42 dB of control field attenuation (80\% probe transmission). The retrieved tripod DSP ($\Psi_{T}^{'}$) has two components, $\Psi_{u}^{'}$ and $\Psi_{d}^{'}$ with a frequency difference $\delta= \omega_{ue} -\omega_{de}$. We find a suitable $\delta$ by choosing a magnetic field B that maximizes the beat note in the retrieved mode.

\subsection{Measurement of 1+1 Dirac Dynamics.}
Once suitable atom-light detunings are chosen, we proceed to create the SSL $\bf{\Psi}$. We use two control fields ($\Omega_{u}^+$ and $\Omega_{d}^+$) co-propagating with the probe $E(z,t)$ and two additional counter-propagating control fields $\Omega_{u}^-$ and $\Omega_{d}^-$ (see Fig. 2c). The created SSL components $\Psi^{\pm}$ are then stored. During storage, the temporal interaction of $\Psi^{+}$ and $\Psi^{-}$ follows the Dirac dynamics outlined by equation \ref{eqDirac}. After storage, these dynamics are mapped onto the SSL components $\Psi^{'\pm}$ by applying the counter-propagating control fields $\Omega_{R}^+$ and $\Omega_{R}^-$ (see Fig. 2d). We detect the optical form of $\Psi^{'\pm}$ ($E^{'\pm}(z,t)$) simultaneously in independent photo-detectors.

We vary the storage time for fixed two-photon detuning, thus changing the interaction time between the SSL components. Each pair of correlated experimental points are obtained by measuring the respective storage of light signals, integrating its total energy for varying storage time. We observe coupled oscillations for the intensities retrieved in each direction, $|E^{'+}(z,t)|^2$ (blue dots) and $|E^{'-}(z,t)|^2$ (red dots) in Fig 3, as expected from the  usual Dirac dynamics coupling the two components of the spinor. Most importantly, the frequency of the oscillation is changed by varying the two-photon detuning, which testifies to the coherent nature of the process (see Fig. 3a and 3b). We note that similar oscillations between SSL frequency components have been shown in previous studies \cite{Lee2014}. However, in our implementation the two spinor components correspond to different propagation directions, which is the key design element for engineering Dirac dynamics.

\begin{figure}
\centerline{\includegraphics[width=1.0\columnwidth]{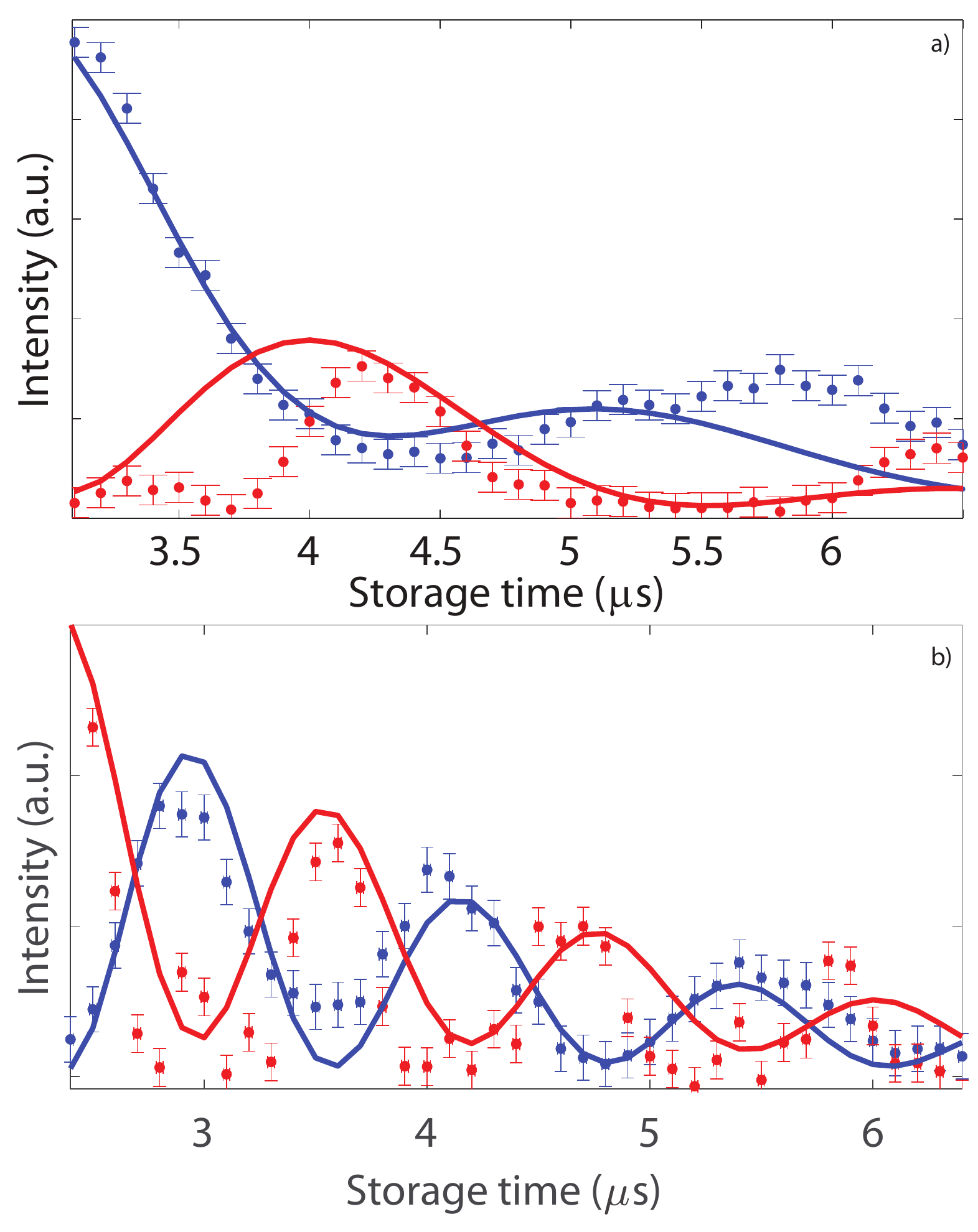}}
\caption{\textbf{ Dirac dynamics using SSL.} Evaluating $|E^{\pm '}|^2$ for each $\tau$ results in an out of phase oscillation between the forward (blue dots) and backward (red dots) components of the SSLs. We plot the experimental data for $\delta$ = 350kHz (c) and 700kHz (d). Solid lines in (c) and (d) corresponds to numerical solutions of the SSL Dirac equation (eq. 2).}
\end{figure}

We benchmarked the aforementioned results against numerical solutions of the 1+1 dimensional Dirac equation of the form (including a coherence decay rate $\gamma$ to account for losses in the real experiment):
\begin{equation*}
i\partial_t \mathbf{\Psi} = \left( iv_g\sigma_z\partial_z + m_{eff_{0}}v_g^2 \sigma_y  -\gamma \right)\mathbf{\Psi},
\end{equation*}
with the initial condition $\mathbf{\Psi_0}=(\Psi_{0}^{+}, \Psi_{0}^{-})$ extracted from the shape of the original SSL right after storage. The solid lines in Fig. 3c and 3d represent the numerical simulation with fixed $v_g=1.0  cm/\mu s$, $m_{eff_{0}}v_g^2=3.3*\delta$ and $\gamma = 0.3$.

As storage time is increased, the SSL components lose their mutual coherence and thus the experimental data begins to deviate from the theoretical prediction. Nonetheless, these measurements provide strong evidence that the SSL dynamics follows that of relativistic particles.

%Topologically Protected Relativistic DynamicsJ
\subsection{Relativistic dynamics with topological behavior.}
Having built an analog Dirac simulator, we now move to mimic JR's topological predictions. In order to engineer $m_{eff}(z)$, we use a spatially varying magnetic field changing the two-photon detuning along the propagation axis of the light. In the original proposal, the bosonic field varies from a negative value to a positive value following the  hyberbolic tangent function \cite{JackiwRebbi1976}. Due to its topological nature however, other profiles exhibiting similar behaviour at the boundary work too, as discussed in detail in \cite{Angelakis2014}. In this work, we choose a linearly-varying magnetic field, $B(z)=B_{0}z$, and perform experiments akin to our previous section. Figure 4a shows the obtained results. We observe the suppression of the coupled oscillation dynamics of the bare Dirac model by increasing the strength of the gradient, as predicted in \cite{JackiwRebbi1976,Angelakis2014}. %The end result for an applied gradient of 435 mG/cm is the full suppression of the oscillations.

We again benchmarked this result against a numerical solution of the Jackiw-Rebbi equation. The procedure is similar to the one presented in the previous section, reconstructing the initial SSL components $\mathbf{\Psi_0}=(\Psi_{0}^{+}, \Psi_{0}^{-})$ (red and blue solid line in fig. 4b) and using $\gamma=\left(\begin{array}{cc} \gamma_{1} & 0\\ 0 & \gamma_{2} \end{array}\right)$ and $m_{eff}(z)v_g^2\propto \delta =(0.745\frac{MHz}{cm}(z-2.5cm) + 0.35MHz)$. In general each of the wave-functions can be written as $\psi^{\pm}(x) = e^{i\Phi^{\pm}(x)}|\psi^{\pm}(x)|$. Assuming $|\Phi^{\pm}(x)|$ to be constant, we define a global phase between the SSL components that is represented as $\Phi$ (a free parameter in the numerical fit). The numerical solutions ($\int dz |\Psi^+(z)|^2$ and $\int dz |\Psi^-(z)|^2$) are obtained numerically by fixing the group velocity, effective mass and loss rate to experimental estimations. The free parameters $\Phi$ and initial relative intensity are then fitted to the data. The solid lines in Fig. 4a (red and blue) represent the numerical solution with $\Phi=\pi$.

To probe the creation of the predicted JR zero energy mode, we first construct $\phi(z)$ (see Fig. 4b, green line) using the effective mass provided by the magnetic field gradient (purple line in Fig. 4b). We then calculate the overlap of the experimentally extracted $\mathbf{\Psi_0}=(\Psi_{0}^{+}, \Psi_{0}^{-})$ (red and blue lines in Fig. 4b, measured at 1.5 $\mu s$ storage time) with the zero-mode spinor: $\int dz \mathbf{\Psi_{0}^{\dagger}} \psi_{zero}$. In Fig. 4c, we plot this overlap for different values of the global phase between the SSL components, $\Phi$. Noticeably, the best overlap of $\sim 80 \%$ is also obtained for $\Phi=\pi$, which strongly hints that $\mathbf{\Psi_0}$ was prepared in the zero-mode.

In our experimental results, we have two decay rates for $\Psi^{+ '}$ and $\Psi^{- '}$ as they couple to a magnetic field insensitive and a magnetic field sensitive EIT line, respectively (see Fig. 2a) \cite{Maynard_Side2015}. Our interpretation of the high initial overlap at 1.5 $\mu s$ storage time is that we have created a zero-energy mode within the medium during this initial time interval.

\begin{figure}
\centerline{\includegraphics[width=1\columnwidth]{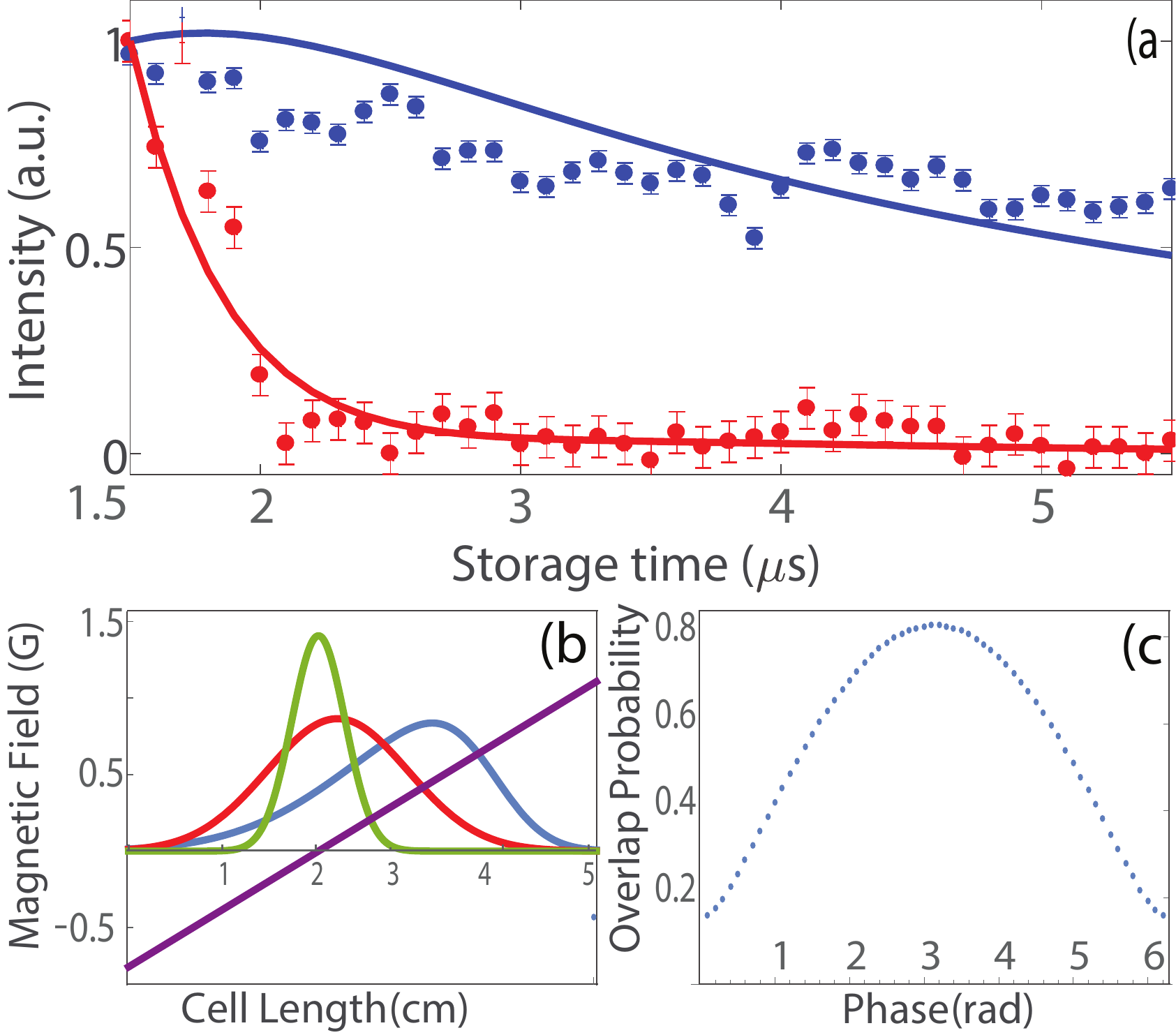}}
\caption{\textbf{JR dynamics using SSL}. (a) Evaluating $|E^{\pm '}|^2$ for each $\tau$ shows a suppression of the oscillation between $E^{+ '}$ (blue dots) and $E^{- '}$ (red dots) as predicted by the JR  seen in Fig. 3c. Full oscillation suppression is obtained for a gradient of 435 mG/cm. (b) The spatial location of initial SSL components $\Psi_{0}^{+}$ (blue) and $\Psi_{0}^{-}$ (red) together with $\phi(z)$ (green) in the cell. Magnetic gradient (purple). (c) $\int dz \mathbf{\Psi_{0}^{\dagger}} \psi_{zero}$ as a function of the phase between $\Psi^{+ '}$ and $\Psi^{- '}$.}
\end{figure}

\section{Discussion}

We have experimentally demonstrated the realization of a controllable coupling between two counter-propagating SSL components, simulating the dynamics of a relativistic massive fermion in a 1+1 Dirac equation. By adding a static background bosonic field (via the use of a magnetic field gradient), we have also simulated the celebrated Jackiw-Rebbi model. We have benchmarked our work with theoretical simulations by carefully reconstructing the initial SSL wave functions and using them in a numerical solution of the corresponding Dirac and JR differential equations. These values are then compared with the experimental data achieved by varying the storage time, showing a very good correlation within the coherence time of the atoms. Lastly, we have also measured signatures of the JR zero-mode by observing the inhibition of the oscillation between the spinor components as predicted in the theory.

We consider our experiment to be an important first step towards more complex quantum simulations with many quantum relativistic particles. Possible extensions include the study of the Klein paradox \cite{Klein1929} or the MIT bag model \cite{Chodos1974_MITbag} by using coupled light-matter SSLs. Moreover, as slow light polaritons can be made to interact strongly, our work provides a pathway towards analog simulators of complex phenomena described by interacting quantum field theories. Possibilities include the simulation of: charge fractionalisation in bosons \cite{Semenoff1982}, the interacting random Dirac model \cite{Keil2013Random} and the renormalization of mass due to interacting fermions \cite{Shankar1994}. Furthermore, interacting relativistic models such as the famous Thirring model \cite{Thirring_1958} are now within experimental reach. As many of these important QFT predictions are only addressable using high energy experiments, this new breed of light-matter room temperature simulators will be an exciting tool to reach unexplored realms of physics.

\section{Acknowledgements}
We thank V. Korepin for insightful discussions. S.B. work was supported by the U.S. Navy  Office of Naval Research, Grant No. N00141410801. B. J. acknowledges financial assistance of the National Research Foundation (NRF) of South Africa. N.U.S work was partially supported by the Singapore Ministry of Education, the Academic Research Fund Tier 3 (Grant No. MOE2012-T3-1-009), the National Research  Foundation (NRF) Singapore and the Ministry of Education, Singapore under the Research Centres of Excellence programme.

%So-found results are compared against the experimental data (see solid lines in Fig. 3a, 3b and 4a).
%We mention that we use only one fitting parameter, related to the initial strength of each spinor components.
%

%\section{References}

%\bibliography{BibtexOct2017}
%\bibliographystyle{prx_with_titles_lastnames}

%\section{Author contribution}
%M. N., B. J. and E. F. performed the experiments and analyzed the data. C. N. and D. A. provided the theoretical and numerical analyses. All authors contributed in the interpretation of the results and the writing of the manuscript.

\end{document}